\RequirePackage{fix-cm}
\documentclass{svjour3}                     
\smartqed  
\usepackage{graphicx}
\usepackage{amsmath}
\usepackage{bm}
\usepackage{subfigure}
\begin{document}

\title{Ground-to-excited derivative couplings for the density
  functional based tight-binding method: Semi-local and long-range
  corrected formulations 
}
\subtitle{}


\author{Thomas A. Niehaus    
}
\dedication{dedicated to Dr.~Fernand Spiegelman on the occasion of his retirement}

\institute{Thomas A. Niehaus \at
               Univ Lyon, Université Claude Bernard Lyon 1, CNRS,
               Institut Lumière Matière, F-69622, Villeurbanne,
               France\\
              Tel.:  +33 472 431 571\\
              \email{thomas.niehaus@univ-lyon1.fr}           
}

\date{Received: date / Accepted: date}

\maketitle

\begin{abstract}
 A derivation of non-adiabatic coupling vectors for the density functional based tight binding method (DFTB) between ground and excited states is presented. The analytical result is valid both for semi-local and long-range corrected DFTB and includes all required Pulay terms. Electron-translation factors lead to a conceptual simplification of the Slater-Koster scheme for precomputed integrals. Compared to scalar  couplings obtained from numerical derivatives, the present approach is computationally more efficient and can be applied to systems with hundreds of atoms. The accuracy of DFTB derivative couplings is assessed by comparison to full density functional theory (DFT) calculations using semi-local and hybrid exchange-correlation functionals with promising results. As exemplified by a case study of furan, DFTB provides non-adiabatic coupling vectors that are close to DFT counterparts in size and direction also in the vicinity of conical intersections.

\keywords{Non-adiabatic coupling \and TD-DFTB \and LC-DFTB \and Furan}
\end{abstract}

\section{Introduction}
\label{intro}
The understanding of radiationless transitions between different electronic quantum states
is key for the further development of optoelectronic devices,
photovoltaics and the optimization of photochemical processes in
general. The probability to perform a transition is driven by
nuclear motion and related to the (first order) derivative coupling 
\begin{equation}
  \label{dIJ}
  d_{IJ}^{\xi} = \langle \Psi_I | \frac{d}{d\xi}| \Psi_J\rangle,
\end{equation} 
between the electronic states $I$ and $J$ for a nuclear coordinate
$\xi$ \cite{Persico2018}.   
Going beyond the Born-Oppenheimer approximation, several
quantum-classical non-adiabatic molecular dynamics (NAMD) methods like the
Ehrenfest approach, Tully's surface hopping (SH) or multiple spawning
techniques have provided useful information on relevant
photochemical pathways on an atomistic scale (for a recent review see \cite{Nelson2020}). Since the potential
energy surfaces (PES) of technologically relevant materials are too
complex for an exhaustive global characterization, on-the-fly NAMD in
which the PES, its gradients and the non-adiabatic coupling are
evaluated for each time step are getting more and more popular. These
electronic structure calculations can be performed at different
quantum chemical levels, with time-dependent density functional
theory (TD-DFT) \cite{Ullrich2012} providing a good compromise between accuracy and
cost. Even though TD-DFT is significantly faster than explicitly
correlated methods for the excited state, the quest for realistic and
functional models involving hundreds of atoms often forces users to limit
their calculations to a handful of trajectories. While these might
provide some insights into possible reaction paths, a converged
determination of properties like for example quantum yields requires a full
sampling of phase space. For the SH approach this is even more
pertinent, since multiple trajectories need to be propagated for each
initial position and momentum pair of the nuclei \cite{Tully1990}. Because of this,
more approximate quantum chemical methods, like the density functional
based tight-binding (DFTB) scheme \cite{Seifert1986,elstner1998scc,frauenheim2002asc}
and its time-dependent extension TD-DFTB \cite{Niehaus2001a,Niehaus2009}, found their niche in NAMD
simulations \cite{Torralva2001,Niehaus2005,Mitric2009,Gao2012a,Stojanovic2017,Bonafe2017,Humeniuk2017,Uratani2020}. Though analytical derivative couplings between ground and excited
states \cite{Chernyak2000,Baer2002a,Tapavicza2007,Hu2007,Hu2008,Tavernelli2009,Hu2009,Send2010} and later also between excited states \cite{Li2014,Ou2015a,Zhang2015,Parker2019a} have been published for
TD-DFT,  no such derivation exists for TD-DFTB. (For attempts to approximate
TD-DFTB couplings see \cite{Humeniuk2019}). This does not impede SH simulations, as most
implementations require only scalar time-derivative couplings  $\sigma(t) = \langle
\Psi_I | \partial/\partial t| \Psi_J\rangle =  \bm{d_{IJ}}
\cdot \bm{\dot{\xi}}$, which can be computed by numerical
differentiation without explicit recourse
to  $\bm{d_{IJ}}$. Such an approach can be quite efficient
\cite{Pittner2009,Alonso-Jorda} so that 
the coupling evaluation is at least for
small molecules not the computational
bottleneck in a NAMD simulation. Questions remain about the validity of numerical couplings
in the context of TD-DFT, since the theory does not provide 
many-body wave functions and one is forced to apply ad-hoc
approximation to this quantity \cite{Tapavicza2007,Werner2008}. A further drawback of scalar
time-dependent couplings is related to the required velocity rescaling
after surface hops to conserve the total energy of ions and
electrons \cite{Coker1995,Herman1984}. Recent investigations indicate that rescaling along the
derivative coupling vector leads to better compliance with detailed
balance and is preferable over other rescaling protocols \cite{Carof2017,Plasser2019}. 

In this article TD-DFTB derivative couplings will be derived following
the TD-DFT approach of Send and Furche for localized atomic
orbital basis sets \cite{Send2010}. Given that general state to state couplings
require higher order response theory and warrant special attention
with regards to the pole structure of the response \cite{Parker2016}, the discussion is
restricted to the simpler case of ground to excited state
couplings. Formulas will be given for both the conventional DFTB
method that is based on semi-local exchange-correlation functionals
in the generalized gradient approximation (GGA), as well as the recent
extension for range-separated functionals, named LC-DFTB \cite{Niehaus2012,Lutsker2015}. Numerical
tests on a selection of small organic molecules will be shown in
section \ref{small}, while section \ref{fur} provides a closer look at the
behavior of TD-(LC)-DFTB  couplings close to a conical intersection (CI).         
                 
\section{Theory}
  DFTB can be derived from a Volterra expansion of the DFT energy
functional around a suitable reference density \cite{Elstner2014}. The most widely
applied model (DFTB2, often also referred to as SCC-DFTB) occurs if this expansion is limited to 
second order. The third order expansion (DFTB3) improves
certain ground state properties \cite{Gaus2011} and can also be extended to excited
states in linear response. It has been shown that results for TD-DFTB2 and TD-DFTB3
differ only slightly \cite{Nishimoto2015} and hence only the former method is described here
in detail. The mentioned DFTB models are all based on local or semi-local
exchange-correlation functionals in the GGA. By contrast, long-range corrected DFTB (LC-DFTB)
involves a fraction of non-local Hartree-Fock exchange. In order to
fix a unified notation for both, the ground and excited state
formulation of DFTB is shortly reviewed.

\subsection{Ground state DFTB}
Long-range corrected (a.k.a. range separated)  DFT methods are characterized by a splitting of the
Coulomb potential ($v_C$) into short-range (sr) and long-range (lr) contributions. In
LC-DFTB this separation is realized using the Yukawa potential 
\begin{equation}
  \label{eq:2}
  v_C = v_C^\text{sr} + v_C^\text{lr} = \frac{\exp(-\omega r_{12})}{r_{12}} +
    \frac{1- \exp(-\omega r_{12})}{r_{12}},
\end{equation}
where $r_{12}$ denotes the electron-electron distance and $\omega$ is
the range-separation parameter. As shown in much more detail in
\cite{Niehaus2012,Lutsker2015}, the LC-DFTB Hamiltonian takes
the following form in the atomic orbital basis ($\sigma, \sigma'$ are spin indices):
\begin{eqnarray}
\label{ham}
  H_{\mu\nu}^\sigma &=& H^{0,w}_{\mu\nu} + \sum_{\sigma'}\sum_{\alpha\beta}
  \Delta D^{\sigma'}_{\alpha\beta} (\mu\nu|v_C + f^\text{xc,w}_{\sigma\sigma'}|\alpha\beta) \nonumber \\
    &&- \frac{c_x}{2} \sum_{\alpha\beta} \Delta D^\sigma_{\alpha\beta} (\mu\alpha|v_C^\text{lr,w}|\beta\nu).
\end{eqnarray}
Here $\bm{H}^{0,w}$ is the LC-DFT Hamiltonian evaluated at a
spin-unpolarized reference
density, $f^\text{xc}({\bf r},{\bf r'})$ denotes the exchange-correlation kernel, and the latter two terms involve two-electron integrals  over
the total and long-range Coulomb potential, respectively. They are
further approximated and simplified in the Mulliken approximation
\cite{Niehaus2012,Lutsker2015}. Eq.~\ref{ham} contains the original
DFTB2 scheme as a special case. The prefactor $c_x$ is used to
discriminate between LC-DFTB ($c_x =1$) and semi-local DFTB 
($c_x =0$). In the latter case, all quantities in eq.~\ref{ham} should
be viewed as independent of the range-separation parameter  $\omega$.  
Solution of the Kohn-Sham equations   
 \begin{equation}
 \sum_{\nu} H^\sigma_{\mu\nu} c^\sigma_{\nu i} = \epsilon_{i\sigma}
 \sum_{\nu} S_{\mu\nu} c^\sigma_{\nu i},
\end{equation}
with $\bm{S}$ being the overlap matrix, allows one to construct the
density matrix $D^\sigma_{\mu\nu}= \sum_i n_{i\sigma} c^\sigma_{\mu i}
c^\sigma_{\nu i}$ and $\bm{\Delta
D}^\sigma = \bm{D}^\sigma -
\bm{D}^{\sigma,0}$. Here $ \bm{D}^{\sigma,0}$ refers to the density
matrix of the reference density and $n_{i\sigma}$ are orbital
occupations. Further, molecular orbital coefficients for orbital $i$ are denoted by
$c^\sigma_{\nu i}$ and the corresponding orbital energies by $\epsilon_{i\sigma}$.  

\subsection{Excited state DFTB: TD-(LC)-DFTB}
Electronic excited states and response properties are available in
TD-DFT and TD-HF through the RPA equations
\cite{Casida1995,Furche2002,Ullrich2012}:

\begin{equation}
  \label{rpa}\left(
  \begin{matrix}
    \bm{A} & \bm{B} \\
    \bm{B} & \bm{A} 
  \end{matrix}
  \right)
  \left(
    \begin{matrix}
      \bm{X} \\\bm{Y} 
    \end{matrix}
  \right) =  \Omega 
   \left(
  \begin{matrix}
    \bm{1} & 0 \\
    0 & \bm{-1}
  \end{matrix}
  \right)
  \left(
    \begin{matrix}
      \bm{X} \\\bm{Y} 
    \end{matrix}
    \right), 
\end{equation}

where the eigenvectors $\bm{X},\bm{Y}$ determine the
transition density and oscillator strength of a certain excited state,
while $\Omega$ denotes the associated transition energy. In the
following we denote
general molecular orbitals (MO) with the indices \{p,q,$\ldots$\}, occupied orbitals
with indices \{i,j,$\ldots$\}, and virtual (unoccupied) orbitals with indices
\{a,b,$\ldots$\}. The matrices
$\bm{A}$ and $\bm{B}$ take the form \cite{Casida1995}:
\begin{eqnarray}
 A_{ia\sigma, jb\sigma'} &=& \frac{\delta_{ij} \delta_{ab} \delta_{\sigma \sigma'}\omega_{jb\sigma'}}{ n_{j\sigma'}-n_{b\sigma'}} + K_{ia\sigma,jb\sigma'}\nonumber\\
 B_{ia\sigma, jb\sigma'} &=&  K_{ia\sigma,bj\sigma'},
\label{ABC}
\end{eqnarray}
where $\omega_{jb\sigma'} = \epsilon_{b\sigma'} - \epsilon_{j\sigma'}$ with $n_{i\sigma} >
n_{a\sigma}$ and $n_{j\sigma'} > n_{b\sigma'}$.

Up to this point the formulation for first principles DFT and DFTB is
identical. For TD-(LC)-DFTB the four-center integrals contained in the
coupling matrices $\bm{K}$ are approximated as follows:

\begin{equation}
  \label{coupling}
K_{ia\sigma,jb\sigma'} = \sum_{Al}\sum_{Bl'} q^{ia\sigma}_{Al}
\Gamma^{\sigma\sigma'}_{Al,Bl'} q^{jb\sigma'}_{Bl'} - c_x\delta_{\sigma\sigma'} q^{ij\sigma}_{Al}
\gamma^{\text{lr}}_{Al,Bl'} q^{ab\sigma'}_{Bl'}.
\end{equation}
The definition of the functions $\Gamma^{\sigma\sigma'}_{Al,Bl'}$ and $\gamma^{\text{lr}}_{Al,Bl'}$, as well as the
transition charges $q^{pq\sigma}_{Al}$ has been given in
\cite{Kranz2017}. The excited state energies and RPA eigenstates are
finally obtained by solving eq.~\ref{rpa}. Numerical algorithms
for this task have been proposed in \cite{Stratmann1998}.

\subsection{Derivative couplings}
As shown by Chernyak and Mukamel \cite{Chernyak2000}, the first-order derivative couplings between the ground and an excited
state $I$
\begin{equation}
  \label{d0I}
  d_{0I}^{\xi} = \langle \Psi_0 | \frac{d}{d\xi}| \Psi_I\rangle,
\end{equation}

can be exactly obtained in TD-DFT through linear response theory,
without explicit construction of a many-body wave function. Send and
Furche \cite{Send2010} derived formulas for the derivative couplings
in a finite atom-centered basis set, that include Pulay type terms
which aid the convergence towards the exact result in the basis set
limit. Their main result holds without changes also for TD-(LC)-DFTB
and reads 
   
\begin{equation}
  \label{d0Ifull}
  d_{0I}^\xi = \sum_{ia\sigma} \left( \frac{1}{\Omega_I}
    (X+Y)_{ia\sigma} R_{ia\sigma}^{(\xi)} - \frac{1}{2}
    (X-Y)_{ia\sigma}  (S_{ia}^{(\xi)} + T_{ia}^{(\xi)})\right). 
\end{equation}
Here $S_{ia}$ denotes a matrix element of the overlap matrix in the
MO basis and the superscript $(\xi)$ indicates a
derivative at fixed MO coefficients. The term $R_{ia}$ is the right
hand side of the coupled perturbed Kohn-Sham (CPKS) equations (see
\cite{Deglmann2002a} and references therein). In
contrast to derivative couplings between excited states, these have
actually never to be solved and the readily available right hand side
is sufficient. In the DFTB context, the term $T_{ia}$ which is obtained from
\begin{equation}
  \label{tia}
  T_{\mu\nu}^{\xi} = \langle \mu |  \nu^\xi\rangle - \langle \mu^\xi |  \nu\rangle,  
\end{equation}
is potentially more problematic. Most DFTB implementations store interatomic Hamiltonian and overlap matrices for varying 
distance as precalculated tables. Slater-Koster rules are used
to construct the molecular Hamiltonian from a small number of high
symmetry integrals and suitable rotations \cite{Slater1954}. Numerical derivatives of the overlap
matrix with respect to position can then be carried out
efficiently. The terms in eq.~\ref{tia} could be evaluated in a similar
fashion, but this would require the implementation of extended Slater-Koster rules. Fortunately, the term involving
$\bm{T}$ can be discarded on physical grounds, which can be most
easily seen in the atomic limit. Here eq.~\ref{d0I} predicts a non-vanishing derivative coupling and
therefore non-adiabatic transitions for an atom moving at constant
velocity. In order to restore translational invariance, so called
electron translation factors (ETFs) can be introduced
\cite{Fatehi2011}. We follow
Ref.~\cite{Parker2019} and neglect the contribution due to $\bm{T}$,
which is responsible for the erroneous atomic limit. 

Similar to the derivation in \cite{Send2010}, eq.~\ref{d0I} may be rewritten in a form that strongly resembles the
formula for TD-(LC)-DFTB  excited state gradients \cite{heringer2007aes,heringer2007aes_erratum,Humeniuk2017}. To this end we
define 
\begin{equation}
  \label{pia}
 \tilde{P}_{ia\sigma} =  \frac{1}{\Omega} (X+Y)_{ia\sigma}, 
\end{equation}
which replaces the relaxed one-particle excited state density matrix for
the gradient evaluation,
\begin{eqnarray}
  \label{wia}
 \tilde{W}_{ia\sigma} &=&  \epsilon_{i\sigma}  \tilde{P}_{ia\sigma} +
  \frac{1}{2} (X-Y)_{ia\sigma} \nonumber\\ 
 \tilde{W}_{ij\sigma} &=& \frac{1}{1+\delta_{ij}} H^+_{ij\sigma}[P],
\end{eqnarray}
which replaces the energy-weighted density matrix, and finally 

\begin{equation}
  \label{gammafr}
  \tilde{\Gamma}_{\mu\nu\sigma, \kappa \lambda \sigma'}  = \tilde{P}_{\mu\nu\sigma}
  \Delta D_{\kappa\lambda\sigma'}, 
\end{equation}
as well as
\begin{eqnarray}
  \label{gammalr}
  \tilde{\Gamma}^\text{lr}_{\mu\nu\sigma, \kappa \lambda \sigma'}  =&&
  -c_x \delta_{\sigma\sigma'} \tilde{P}_{\mu\lambda\sigma}
  \Delta D_{\nu\kappa\sigma},
\end{eqnarray}
which can be seen as replacements for the two-particle excited state
density matrix. Apart from the last two equations, these DFTB expressions are identical to their first
principles counterparts. The fact that $\bm{
  \tilde{\Gamma}}$ differs from TD-DFT is because the DFTB
zero-order Hamiltonian $\bm{H^0}$ already includes the Hartree and
exchange-correlation potentials of the reference density. Consistent with the typical TD-(LC)-DFTB
two-center Mulliken approximations, one further has   
 
\begin{eqnarray}
  \label{H+-}
H^+_{pq\sigma}[V] = && \sum_{rs\sigma'} \sum_{Al} \sum_{Bl'}\left\{ 2 q^{pq\sigma}_{Al}
                       \Gamma^{\sigma\sigma'}_{Al,Bl'}
                       q^{rs\sigma'}_{Bl'} \right.\nonumber \\
  && - c_x \left. \delta_{\sigma\sigma'} \left[ q^{ps\sigma}_{Al}
     \gamma^\text{lr}_{Al,Bl'} q^{rq\sigma}_{Bl'} + q^{pr\sigma}_{Al}
     \gamma^\text{lr}_{Al,Bl'} q^{sq\sigma}_{Bl'} \right] \right\}V_{rs\sigma'}.
 \end{eqnarray}
for a general vector $\bm{V}$. After transformation to the atomic
orbital basis, the main result finally reads:
\begin{eqnarray}
  \label{forces}
 d_{0I}^\xi  =&& \sum_{\mu\nu \sigma}\left(
  H^{0,\xi}_{\mu\nu} \tilde{P}_{\mu\nu\sigma} -
  S_{\mu\nu}^{\xi} \tilde{W}_{\mu\nu\sigma}\right) \nonumber\\
&& + \sum_{\mu\nu\sigma\kappa\lambda\sigma'}   (\mu\nu|v_C +
   f^\text{xc,w}_{\sigma\sigma'}|\kappa\lambda)^{\xi}
   \tilde{\Gamma}_{\mu\nu\sigma, \kappa \lambda \sigma'}  \nonumber\\
&& + \sum_{\mu\nu\sigma\kappa\lambda\sigma'}  (\mu\nu|v_C^{\text{lr},\omega} 
   |\kappa\lambda)^{\xi}
   \tilde{\Gamma}^\text{lr}_{\mu\nu\sigma, \kappa \lambda \sigma'}, 
\end{eqnarray}

where the remaining derivatives of the two-electron integrals in the two last lines of
eq.~\ref{forces} can be simplified as in \cite{heringer2007aes,heringer2007aes_erratum}.  
The implementation of these equations in the {\tt dftb+} code \cite{Hourahine2020} is currently
limited to singlet excited states. As a useful check, it was verified that the sum of the cartesian components of the derivative coupling vector over all atoms equals zero \cite{Tommasini2001}. A similar sum rule holds for the excited state forces. From the numerical point of view,
the computation of the derivative couplings can be combined with the
evaluation of excited state gradients at negligible additional effort.   
\section{Numerical results}
\subsection{Small molecules at the equilibrium geometry}
\label{small}
In order to test the reliability and accuracy of the TD-DFTB
derivative couplings, first principles TD-DFT calculations 
with the TURBOMOLE version 7.4.1 software package were performed. This
allows for a meaningful comparison, since ETFs are taken into account
in the same way as in DFTB, if the keyword {\tt \$nacme etf} is chosen in
the TD-DFT excited state calculations.  Simulations have
been performed with the semi-local PBE functional, which is the one
used in the generation of Slater-Koster files for the conventional
DFTB method. Specifically, the {\tt mio-0-1} set \cite{elstner1998scc,Niehaus2001} has been employed in the TD-DFTB calculations.  Since range-separated functionals are currently not
available in TURBOMOLE, direct comparison to LC-DFTB results was not
possible. Computations using the hybrid functional B3LYP
have been done instead to estimate the influence of non-local
Hartree-Fock exchange in the functional. TD-LC-DFTB calculations were done with a range-separation parameter of $\omega = $ 0.3 $a_0^{-1}$ as described in \cite{Kranz2017}.

All first principles
calculations are performed using the def2-TZVP basis set. A small
number of molecules from the Schreiber test set for electronic excited
states has been selected, which had been optimized at the MP2/6-31G*
level and are available in the Supplemental Material of that
article. Though derivative couplings at the equilibrium position of
the ground state are typically not relevant for fast photochemical
reactions, they are investigated here to benchmark the absolute value
and direction of the TD-(LC)-DFTB couplings. To this end the norm of
the derivative coupling $|\bm{d}|$ and the angle $\theta$ with
respect to TD-PBE couplings, defined as 
\begin{gather}
  \label{theta}
  \cos{\theta} = \frac{\bm{d}^X \cdot \bm{d}^\text{PBE}}{|\bm{d}^X|  |\bm{d}^{PBE}|}
\end{gather}
for a given method $X$, are listed in Table \ref{tab:ex}. The investigated states are typically the lowest excited singlet states ($S_1$), since these are the most relevant ones for radiationless decay to the ground state, though for Uracil also some higher states are included. The energetical ordering is not the same in all methods, such that the comparison is based on the symmetry of the state, as usual.

Comparing first the two first principles TD-DFT methods, one finds that the magnitude of the derivative couplings for both functionals differs by roughly 9 \% on average. There is no clear trend as to whether PBE or B3LYP provides larger couplings, although B3LYP delivers on average slightly larger excitation energies (see also \cite{Jacquemin2009a} for a full benchmark on this topic) which should lead to smaller couplings according to eq.~\ref{pia}. This illustrates that the precise form of the transition density influences the result to an important degree.  The angle between non-adiabatic coupling vectors for both methods are rather small. The B3LYP result for the second $A''$ state with $\theta =$ 35.7$^\circ$ is an exception. Here the application of the measure defined in eq.~\ref{theta} leads to a perhaps counter-intuitive result. Although the coupling vector on each atom is exactly parallel for B3LYP and PBE (namely perpendicular to the molecular plane) a non vanishing angle arises. The chosen measure is still relevant, since the non-adiabatic transition probability is given by the scalar product of derivative coupling and the $3N$ dimensional velocity vector of the ions, $N$ being the number of atoms. Comparing now TD-PBE and TD-DFTB, which utilize the same exchange-correlation functional, there is again no clear trend in the deviations, which are on average  13 \% for the coupling magnitude, with a maximum deviation of 22 \% for benzene. Angles show likewise larger deviations as the TD-B3LYP results. Given the rather good agreement of  TD-PBE and TD-DFTB excitation energies, the difference can be attributed to the minimal basis set in TD-DFTB which should lead to a less accurate representation of the transition density. The TD-LC-DFTB results show a smaller deviation with respect to TD-B3LYP than to TD-PBE. In addition, TD-LC-DFTB couplings are systematically larger (smaller) than TD-DFTB couplings, in cases where TD-B3LYP couplings are larger (smaller) than TD-PBE couplings. This indicates that the nature of the functional (semi-local vs. non-local) is still visible in the DFTB results, in spite of the various approximations in going from DFT to DFTB. Though the present survey is far from being exhaustive, one can state a reasonable agreement between TD-DFT and TD-DFTB results.            

\begin{table}
\caption{\label{tab:ex} Ground-to-excited derivative couplings for selected singlet excited states at the TD-DFT level with the PBE and B3LYP functionals together with TD-DFTB and TD-LC-DFTB data. The excitation energy is labeled $\Delta E$ (in eV), norm $|\bm{d}|$ (in $a_0^{-1}$) and angle $\theta$ (in degree) with respect to PBE couplings are defined in the text.}
    \begin{tabular}{lcccccccccccc}
     Molecule  & State & \multicolumn{2}{c}{PBE}  & \multicolumn{3}{c}{B3LYP} & \multicolumn{3}{c}{DFTB} & \multicolumn{3}{c}{LC-DFTB} \\
               &       & $\Delta E$ & $|\bm{d}|$ & $\Delta E$ & $|\bm{d}|$ & $\theta$ & $\Delta E$ & $|\bm{d}|$ & $\theta$ & $\Delta E$ & $|\bm{d}|$ & $\theta$   \\\hline
 Ethene & $B_{1u}$  & 7.75 & 0.049 & 7.69 & 0.037 & 7.1  & 7.67 & 0.059 & 6.8  & 8.36 & 0.045 &  9.5 \\
 Butadiene & $B_u$ & 5.60 & 0.110 & 5.72 & 0.117 & 3.8  & 5.50 & 0.091 & 28.0 & 6.17 & 0.088 & 22.8 \\
 Benzene & $B_{2u}$ & 5.25 & 1.218 & 5.39 & 1.287 & 0.1  & 5.30 & 1.483 & 1.1  & 6.07 & 1.560 & 1.1 \\
 Formamide & $A''$ & 5.43 & 0.288 & 5.55 & 0.287 & 1.0  & 5.51 & 0.316 & 7.1  & 5.89 & 0.329 & 6.3 \\
 Furan & $B_2$     & 6.09 & 0.622 & 6.12 & 0.659 & 0.8  & 6.06 & 0.693 & 7.0  & 6.61 & 0.696 & 8.3 \\
 Uracil & $A''$    & 3.92 & 0.242 & 4.60 & 0.273 & 9.4  & 3.70 & 0.256 & 22.3 & 4.71 & 0.288 & 23.3 \\
 & $A''$           & 4.71 & 0.212 & 5.72 & 0.202 & 35.7 & 4.24 & 0.198 & 18.7 & 5.80 & 0.206 & 25.3 
 \\& $A'$          & 4.74 & 0.759 & 5.15 & 0.858 & 5.1  & 5.15 & 0.673 & 28.0 & 5.75 & 0.991 & 17.2
    \end{tabular}
\end{table}
\subsection{Behavior close to conical intersections: The case of furan}
\label{fur}
\begin{figure}[htbp]
  \subfigure[][]{	\includegraphics[width=0.3\textwidth]{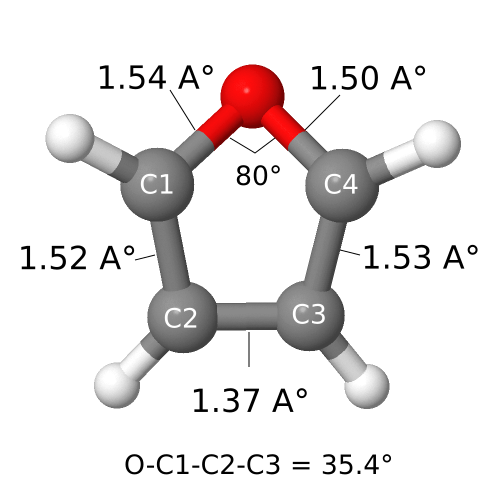}	}
\hfill
  \subfigure[][]{ \includegraphics[width=0.3\textwidth] {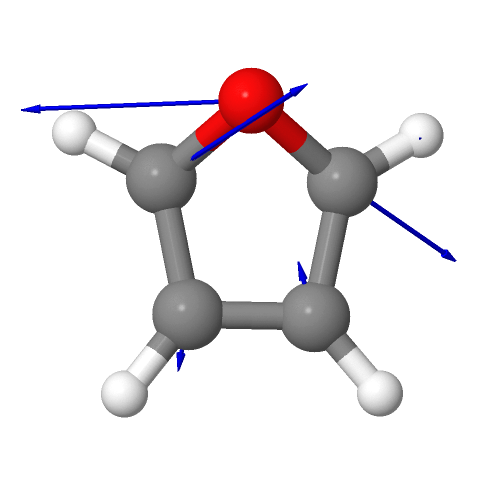}}
\hfill
 \subfigure[][]{\includegraphics[width=0.3\textwidth] {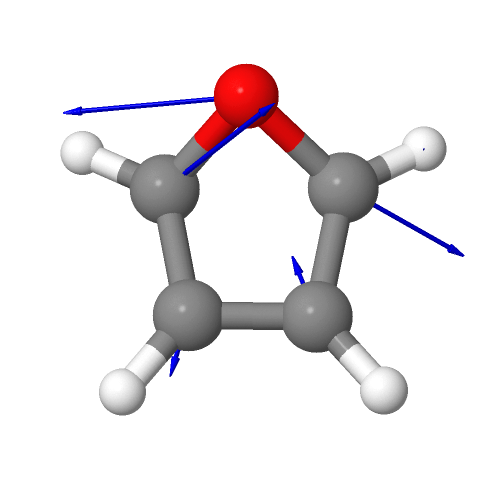}}
  \caption{\label{cifig} (a) Furan geometry at the approximate
    S$_1$/S$_0$ conical intersection for the TD-DFTB method.  The
    dihedral angle of oxygen with the carbon ring is given at the
    bottom.  (b) TD-DFTB derivative coupling vector $\bm{d}_{01}$ (blue
    arrows) at point 17 of the reaction path mentioned in the
    text. (c)   TD-PBE derivative coupling vector at the same geometry.} 
\end{figure}
As a further test of the implementation, it is useful to study the
couplings close to a conical intersection where the potential energy
surfaces (PES) of ground and excited states become degenerate. At such
crossing seams of the PES the coupling will diverge, but even at
geometries with a finite, but small gap, the derivative couplings are
large such that surface crossings become highly probable. The target
in this section is furan, a heterocyclic ring molecule that was
recently experimentally studied as a model system in attosecond
spectroscopy to unravel the ultrafast motion through a CI \cite{Liu2015,Hua2016,Adachi2019}. In the
past, furan has been extensively investigated with a large variety of
quantum chemical methods \cite{Serrano-Andres1993,Palmer1995,Burcl2002,Gromov2003,Gavrilov2008,Stenrup2011}. Its two lowest lying singlet excited states
are given by a state of Rydberg character ($A_2(3s)$, exp: 5.94
eV, 5.91 eV) and a valence excited state of $\pi \to
\pi^*$ type ($B_2(V)$, exp: 6.06 eV, 6.04 eV) \cite{Serrano-Andres1993,Palmer1995}. Since DFTB
employs a minimal basis, only the valence excited state can be
resolved and is found at 6.06 eV (TD-DFTB) and 6.61 eV
(TD-LC-DFTB). Using MS-CASPT2 theory, Stenrup and Larsen \cite{Stenrup2011} were
able to locate two low energy CI corresponding to a ring opened and
ring puckered geometry. The latter is related to the degeneracy of the
$\pi \to \pi^*$ state with the ground state and studied in greater
detail in this section. In order to explore the PES, the CI optimizer
by Bearpark and co-workers was implemented \cite{Bearpark1994}. Their method
minimizes the gradient difference vector between two states $\bm{g}$
together with the projection of the upper state gradient on the
intersection plane spanned by $\bm{g}$ and $\bm{d}$. As also mentioned
by the authors, this algorithm leads to a quick reduction of the
energy gap due to minimization of $\bm{g}$, eventually leading to self
consistent field (SCF) convergence problems, while the projection on
the intersection plane remains sizable. To avoid such complications, a
recent extension of the Bearpark algorithm was also tested
\cite{Harabuchi2019}, which showed no significant improvement for the
present method and system. It should be noted  that TD-DFT provides a
wrong topology of S$_1$/S$_0$ intersections \cite{Levine2006a} and
TD-DFTB as a single reference method shares this shortcoming. This
does however not preclude the localization of PES intersections.

\begin{figure}[htbp]
  \includegraphics[width=\textwidth]{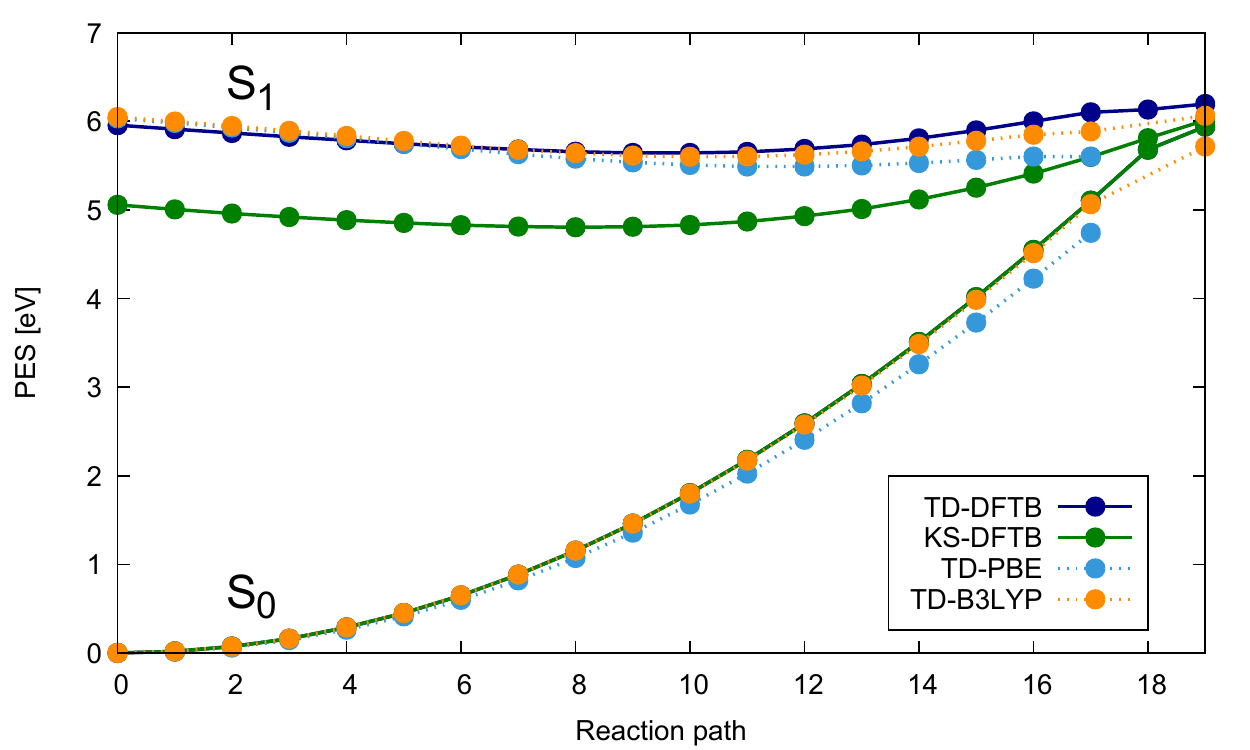}
  \caption{\label{pes} Furan S$_0$ and S$_1$ PES at different
    levels of theory. The reaction path is obtained from a linear
    interpolation between the DFTB S$_0$ minimum (value 0) and the
    TD-DFTB approximate CI shown in Fig.~\ref{cifig} (value 19). Lines
    are guides to the eyes. Missing data points indicate SCF failure. The
    S$_0$ PES for DFTB is identical to KS-DFTB (see text). Total energies for
    LC-DFTB are not shown, because the required repulsive
    potentials were not available. 
} 
\end{figure}
\begin{figure}[htbp]
  \includegraphics[width=\textwidth]{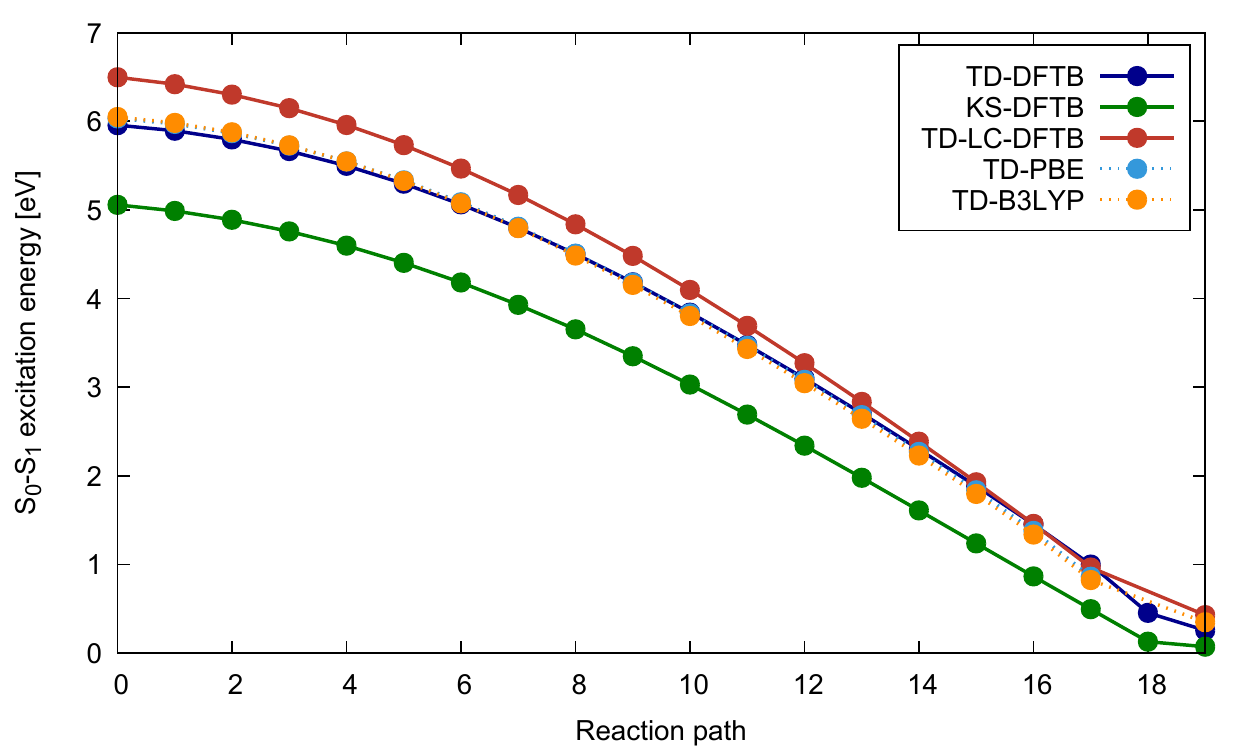}
  \caption{\label{exc}  S$_1$-S$_0$ gap along the
    reaction path.  
} 
\end{figure}

Starting from the ground state equilibrium geometry of furan, the CI
optimizer was used to find the closest S$_1$/S$_0$ intersection from
the Franck-Condon point for the TD-DFTB method. Fig.~\ref{cifig} shows
the last geometry along the optimization path, before the
self-consistent charge loop in DFTB failed to converge. At this point
the TD-DFTB S$_1$/S$_0$ gap is 0.254 eV. This approximate CI geometry
exhibits two O-C bonds of similar length, while the ring puckered CI
in \cite{Stenrup2011} features a larger difference in bond lengths of 0.3\
\AA. CI optimization starting from the latter structure resulted again
in the structure shown in Fig.~\ref{cifig}. The evolution of the
ground and excited PES going from the S$_0$ minimum to the TD-DFTB CI
is depicted in Fig.~\ref{pes} along an approximate reaction
pathway. For some geometries close to the CI, TD-PBE and TD-B3LYP
could not be converged, which indicates that also for these methods
the TD-DFTB CI presents a structure close to degeneracy. In line with
the results of Stenrup and Larsen \cite{Stenrup2011}, TD-B3LYP predicts
that the ring puckered CI may be reached by a barrier less path from
the Franck-Condon (FC) point ($\Delta E =
E_{S_1}(FC)-E_{S_1}(CI)=0.02$ eV). TD-DFTB provides a value of $\Delta
E = -0.238$ eV, but a generally quite similar PES. It is noted that
the MS-CASPT2 result in \cite{Stenrup2011} is $\Delta E = 0.98$ eV for the
slightly different CI structure mentioned above. 

\begin{figure}[htbp]
  \includegraphics[width=\textwidth]{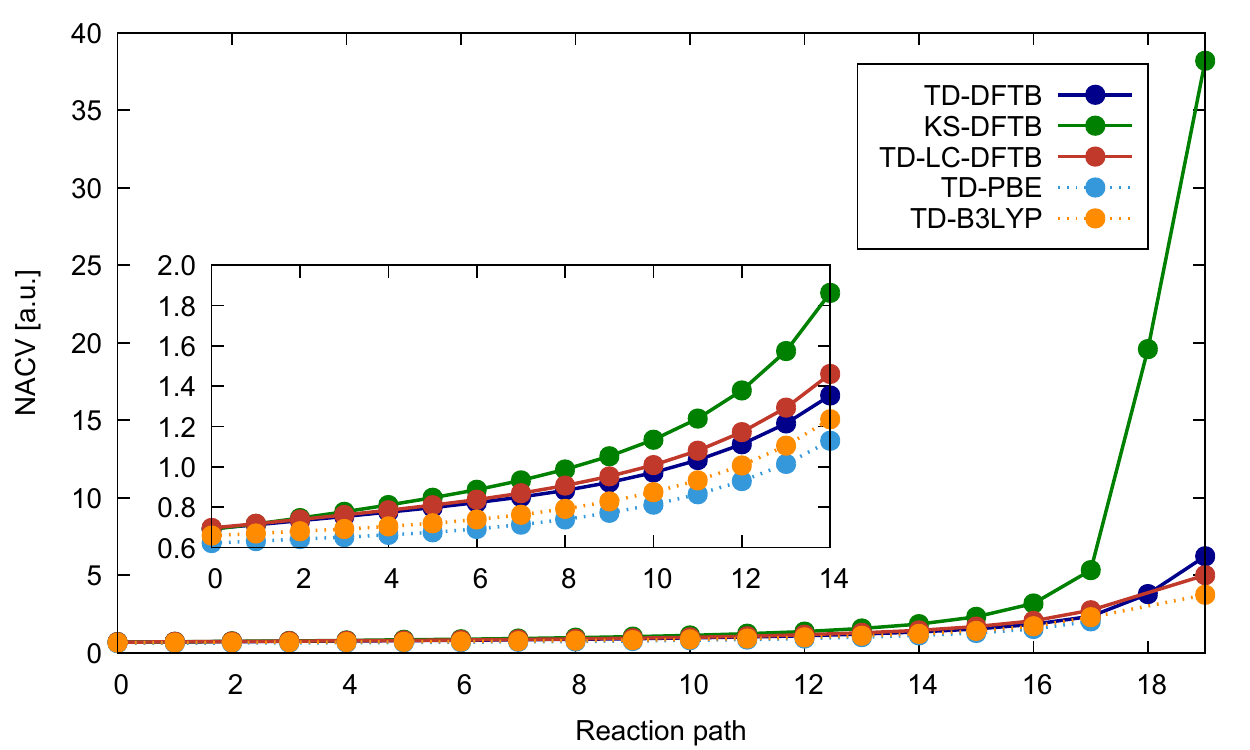}
  \caption{\label{nac}  Norm of the derivative coupling (atomic units)
    along the reaction path. The inset shows the same quantity for a
    smaller range.  
} 
\end{figure}

Discussing now the
derivative couplings along the reaction path depicted in Fig.~\ref{nac},
one observes a steep increase as one approaches the CI. This increase
is related to the decrease of the  S$_1$/S$_0$ gap shown in
Fig.~\ref{exc}. TD-DFTB and TD-LC-DFTB systematically overestimate the
couplings compared to the corresponding
first principles values for TD-PBE and TD-B3LYP, respectively, by
around 10-20 \%. As can be seen in Fig.~\ref{cifig} not only the norm of
$\bm{d}$ but also its direction are in rather good agreement between
DFTB and DFT.  While the SCF for PBE does not converge at the TD-DFTB CI
geometry, there is no such problem for B3LYP. With respect to DFTB,
both B3LYP and LC-DFTB show an inversion of HOMO and LUMO, which explains the reversed TD-LC-DFTB/TD-DFTB ordering of
$|\bm{d}|$ at this point. 

Fig.~\ref{pes} to Fig.~\ref{nac} also include results for the case where the
excited state wave function is represented by a single Kohn-Sham
determinant (KS-DFTB). At the TD-DFT level, this approximation has been frequently used in
the past to reduce the numerical effort and simplify calculations \cite{Tapavicza2007,Meng2008,Meng2008a}.
In this case the excitation energy corresponds to the
HOMO-LUMO MO energy gap and in principle the RPA equations do not need
to be solved. In the present calculations this limit is technically realized by
setting $\Gamma$ and $\gamma^\text{lr}$ to zero
(c.f. eq.~\ref{coupling}), while keeping the rest of the code
unchanged. As Fig.~\ref{nac} shows, this approach leads to grossly
overestimated coupling values close to the CI. In the context of
non-adiabatic MD simulations one should expect strongly overestimated crossing probabilities and
underestimated excited state lifetimes as a result.   

\section{Summary}
The present article summarized the derivation of ground-to-excited derivative couplings for the DFTB and LC-DFTB methods. Existing implementations for excited state gradients can be easily extended to provide also the couplings at very little additional computational cost. In comparison to numerical derivative couplings they are significantly faster: At each geometry only one instead of $3N$ TD-DFTB calculations are required. In addition, numerical approaches require the computation of Slater determinant overlaps. The number of these grows strongly with the size of the system. So far, the accuracy of DFTB derivative couplings has never been investigated. Sec.~\ref{small} indicates a typical deviation of 10 \%, which is acceptable given the larger number of trajectories that can be treated at this level of theory. In the case study of furan it was found that the qualitative agreement with first principles DFT also holds close to conical intersection, which is key for a correct determination of photochemical pathways. The implementation of a CI optimizer using the derivative coupling vector as important ingredient should be useful in further characterizations of DFTB potential energy surfaces. As a natural next step, DFTB couplings between arbitrary excited states need to be developed, which would pave the way for large scale non-adiabatic molecular dynamics simulations. This project is currently underway.              

\begin{acknowledgements}
 Financial support by the Laboratoire d’Excellence iMUST is gratefully acknowledged.
\end{acknowledgements}

%
%


\bibliographystyle{spphys} 
\bibliography{Combined}

\begin{thebibliography}{10}
\providecommand{\url}[1]{{#1}}
\providecommand{\urlprefix}{URL }
\expandafter\ifx\csname urlstyle\endcsname\relax
  \providecommand{\doi}[1]{DOI \discretionary{}{}{}#1}\else
  \providecommand{\doi}{DOI \discretionary{}{}{}\begingroup
  \urlstyle{rm}\Url}\fi

\bibitem{Persico2018}
M.~Persico, G.~Granucci, \emph{Photochemistry: A Modern Theoretical
  Perspective} (Springer, 2018)

\bibitem{Nelson2020}
T.R. Nelson, A.J. White, J.A. Bjorgaard, A.E. Sifain, Y.~Zhang, B.~Nebgen,
  S.~Fernandez-Alberti, D.~Mozyrsky, A.E. Roitberg, S.~Tretiak, Chem. Rev.
  \textbf{120}(4), 2215 (2020)

\bibitem{Ullrich2012}
C.~Ullrich, \emph{Time-Dependent Density-Functional Theory: Concepts and
  Applications} (Oxford University Press, USA, 2012)

\bibitem{Tully1990}
J.C. Tully, J. Chem. Phys. \textbf{93}(2), 1061 (1990)

\bibitem{Seifert1986}
G.~Seifert, H.~Eschrig, W.~Bieger, Z. Phys. Chem. (Leipzig) \textbf{267}, 529
  (1986)

\bibitem{elstner1998scc}
M.~Elstner, D.~Porezag, G.~Jungnickel, J.~Elsner, M.~Haugk, T.~Frauenheim,
  S.~Suhai, G.~Seifert, Phys. Rev. B \textbf{58}(11), 7260 (1998)

\bibitem{frauenheim2002asc}
T.~Frauenheim, G.~Seifert, M.~Elstner, T.~Niehaus, C.~K\"{o}hler, M.~Amkreutz,
  M.~Sternberg, Z.~Hajnal, A.~Di~Carlo, S.~Suhai, J. Phys.: Cond. Mat.
  \textbf{14}(11), 3015 (2002)

\bibitem{Niehaus2001a}
T.A. Niehaus, S.~Suhai, F.~Della~Sala, P.~Lugli, M.~Elstner, G.~Seifert,
  T.~Frauenheim, Phys. Rev. B \textbf{63}(8), 085108 (2001)

\bibitem{Niehaus2009}
T.A. Niehaus, J. Mol. Struct.: THEOCHEM \textbf{914}, 38 (2009)

\bibitem{Torralva2001}
B.~Torralva, T.A. Niehaus, M.~Elstner, S.~Suhai, T.~Frauenheim, R.E. Allen,
  Phys Rev B \textbf{6415}(15), 153105 (2001)

\bibitem{Niehaus2005}
T.A. Niehaus, D.~Heringer, B.~Torralva, T.~Frauenheim, Eur. Phys. J. D
  \textbf{35}(3), 467 (2005)

\bibitem{Mitric2009}
R.~Mitric, U.~Werner, M.~Wohlgemuth, G.~Seifert, V.~Bonacic-Kouteck{\`y}, J.
  Phys. Chem. A \textbf{113}, 12700 (2009)

\bibitem{Gao2012a}
X.~Gao, Q.~Peng, Y.~Niu, D.~Wang, Z.~Shuai, Phys. Chem. Chem. Phys.
  \textbf{14}(41), 14207 (2012)

\bibitem{Stojanovic2017}
L.~Stojanovic, S.G. Aziz, R.H. Hilal, F.~Plasser, T.A. Niehaus, M.~Barbatti, J.
  Chem. Theory Comput. \textbf{13}(12), 5846 (2017)

\bibitem{Bonafe2017}
F.P. Bonaf{\'e}, B.~Aradi, M.~Guan, O.A. Douglas-Gallardo, C.~Lian, S.~Meng,
  T.~Frauenheim, C.G. S{\'a}nchez, Nanoscale \textbf{9}(34), 12391 (2017)

\bibitem{Humeniuk2017}
A.~Humeniuk, R.~Mitric, Comput. Phys. Commun. \textbf{221}, 174 (2017)

\bibitem{Uratani2020}
H.~Uratani, H.~Nakai, J. Chem. Phys. \textbf{152}(22), 224109 (2020)

\bibitem{Chernyak2000}
V.~Chernyak, S.~Mukamel, J. Chem. Phys. \textbf{112}(8), 3572 (2000)

\bibitem{Baer2002a}
R.~Baer, Chem. Phys. Lett. \textbf{364}(1-2), 75 (2002)

\bibitem{Tapavicza2007}
E.~Tapavicza, I.~Tavernelli, U.~Rothlisberger, Phys. Rev. Lett. \textbf{98}(2),
  023001 (2007)

\bibitem{Hu2007}
C.~Hu, H.~Hirai, O.~Sugino, J. Chem. Phys. \textbf{127}(6), 064103 (2007)

\bibitem{Hu2008}
C.~Hu, H.~Hirai, O.~Sugino, J. Chem. Phys. \textbf{128}(15), 154111 (2008)

\bibitem{Tavernelli2009}
I.~Tavernelli, E.~Tapavicza, U.~Rothlisberger, J. Chem. Phys. \textbf{130}(12),
  124107 (2009)

\bibitem{Hu2009}
C.~Hu, O.~Sugino, Y.~Tateyama, J. Chem. Phys. \textbf{131}(11), 114101 (2009)

\bibitem{Send2010}
R.~Send, F.~Furche, J. Chem. Phys. \textbf{132}(4), 044107 (2010)

\bibitem{Li2014}
Z.~Li, B.~Suo, W.~Liu, J. Chem. Phys. \textbf{141}(24), 244105 (2014)

\bibitem{Ou2015a}
Q.~Ou, G.D. Bellchambers, F.~Furche, J.E. Subotnik, J. Chem. Phys.
  \textbf{142}(6), 064114 (2015)

\bibitem{Zhang2015}
X.~Zhang, J.M. Herbert, J. Chem. Phys. \textbf{142}(6), 064109 (2015)

\bibitem{Parker2019a}
S.M. Parker, S.~Roy, F.~Furche, Phys. Chem. Chem. Phys. \textbf{21}(35), 18999
  (2019)

\bibitem{Humeniuk2019}
A.~{Humeniuk}, R.~{Mitri{\'c}}, arXiv e-prints arXiv:1908.00276 (2019)

\bibitem{Pittner2009}
J.~Pittner, H.~Lischka, M.~Barbatti, Chem. Phys. \textbf{356}(1), 147 (2009)

\bibitem{Alonso-Jorda}
P.~Alonso-Jord{\'a}, D.~Davidovi{\'c}, M.~Sapunar, J.R. Herrero, E.S.
  Quintana-Ort{\'\i}, Comput. Phys. Commun. \textbf{258}, 107521 (2021)

\bibitem{Werner2008}
U.~Werner, R.~Mitri{\'c}, T.~Suzuki, V.~Bona{\v{c}}i{\'c}-Kouteck{\`y}, Chem.
  Phys. \textbf{349}(1-3), 319 (2008)

\bibitem{Coker1995}
D.F. Coker, L.~Xiao, J. Chem. Phys. \textbf{102}(1), 496 (1995)

\bibitem{Herman1984}
M.F. Herman, J. Chem. Phys. \textbf{81}(2), 754 (1984)

\bibitem{Carof2017}
A.~Carof, S.~Giannini, J.~Blumberger, J. Chem. Phys. \textbf{147}(21), 214113
  (2017)

\bibitem{Plasser2019}
F.~Plasser, S.~Mai, M.~Fumanal, E.~Gindensperger, C.~Daniel, L.~Gonz{\'a}lez,
  J. Chem. Theory Comput. \textbf{15}(9), 5031 (2019)

\bibitem{Parker2016}
S.M. Parker, S.~Roy, F.~Furche, J. Chem. Phys. \textbf{145}(13), 134105 (2016)

\bibitem{Niehaus2012}
T.~Niehaus, F.~Della~Sala, physica status solidi (b) \textbf{249}, 237 (2012)

\bibitem{Lutsker2015}
V.~Lutsker, B.~Aradi, T.A. Niehaus, J. Chem. Phys. \textbf{143}(18), 184107
  (2015)

\bibitem{Elstner2014}
M.~Elstner, G.~Seifert, Philos. Trans. R. Soc. A \textbf{372}(2011), 20120483
  (2014)

\bibitem{Gaus2011}
M.~Gaus, Q.~Cui, M.~Elstner, J. Chem. Theory Comput \textbf{7}(4), 931 (2011)

\bibitem{Nishimoto2015}
Y.~Nishimoto, J. Chem. Phys. \textbf{143}(9), 094108 (2015)

\bibitem{Casida1995}
M.E. Casida, \emph{Recent Advances in Density Functional Methods, Part I}
  (World Scientific, Singapore, 1995), chap. Time-dependent Density Functional
  Response Theory for Molecules, pp. 155--192

\bibitem{Furche2002}
F.~Furche, R.~Ahlrichs, J. Chem. Phys. \textbf{117}(16), 7433 (2002)

\bibitem{Kranz2017}
J.J. Kranz, M.~Elstner, B.~Aradi, T.~Frauenheim, V.~Lutsker, A.D. Garcia, T.A.
  Niehaus, J. Chem. Theory Comput. \textbf{13}(4), 1737 (2017)

\bibitem{Stratmann1998}
R.E. Stratmann, G.E. Scuseria, M.J. Frisch, J. Chem. Phys. \textbf{109}(19),
  8218 (1998)

\bibitem{Deglmann2002a}
P.~Deglmann, F.~Furche, R.~Ahlrichs, Chem. Phys. Lett. \textbf{362}(5-6), 511
  (2002)

\bibitem{Slater1954}
J.C. Slater, G.F. Koster, Phys. Rev. \textbf{94}(6), 1498 (1954)

\bibitem{Fatehi2011}
S.~Fatehi, E.~Alguire, Y.~Shao, J.E. Subotnik, J. Chem. Phys. \textbf{135}(23),
  234105 (2011)

\bibitem{Parker2019}
S.M. Parker, S.~Roy, F.~Furche, Phys. Chem. Chem. Phys. \textbf{21}(35), 18999
  (2019)

\bibitem{heringer2007aes}
D.~Heringer, T.A. Niehaus, M.~Wanko, T.~Frauenheim, J. Comp. Chem.
  \textbf{28}(16), 2589 (2007)

\bibitem{heringer2007aes_erratum}
D.~Heringer, T.A. Niehaus, M.~Wanko, T.~Frauenheim, J. Comp. Chem.
  \textbf{33}(16), 593 (2012)

\bibitem{Hourahine2020}
B.~Hourahine, B.~Aradi, V.~Blum, F.~Bonafé, A.~Buccheri, C.~Camacho,
  C.~Cevallos, M.Y. Deshaye, T.~Dumitrică, A.~Dominguez, S.~Ehlert,
  M.~Elstner, T.~van~der Heide, J.~Hermann, S.~Irle, J.J. Kranz, C.~Köhler,
  T.~Kowalczyk, T.~Kubař, I.S. Lee, V.~Lutsker, R.J. Maurer, S.K. Min,
  I.~Mitchell, C.~Negre, T.A. Niehaus, A.M.N. Niklasson, A.J. Page, A.~Pecchia,
  G.~Penazzi, M.P. Persson, J.~Řezáč, C.G. Sánchez, M.~Sternberg,
  M.~Stöhr, F.~Stuckenberg, A.~Tkatchenko, V.W.z. Yu, T.~Frauenheim, J Chem
  Phys \textbf{152}(12), 124101 (2020)

\bibitem{Tommasini2001}
M.~Tommasini, V.~Chernyak, S.~Mukamel, Int. J. Quantum Chem. \textbf{85}(4-5),
  225 (2001)

\bibitem{Niehaus2001}
T.A. Niehaus, M.~Elstner, T.~Frauenheim, S.~Suhai, J. Mol. Struct. - Theochem
  \textbf{541}, 185 (2001)

\bibitem{Jacquemin2009a}
D.~Jacquemin, V.~Wathelet, E.A. Perpete, C.~Adamo, J. Chem. Theory Comput.
  \textbf{5}(9), 2420 (2009)

\bibitem{Liu2015}
Y.~Liu, G.~Knopp, C.~Qin, T.~Gerber, Chem. Phys. \textbf{446}, 142 (2015)

\bibitem{Hua2016}
W.~Hua, S.~Oesterling, J.D. Biggs, Y.~Zhang, H.~Ando, R.~de~Vivie-Riedle, B.P.
  Fingerhut, S.~Mukamel, Structural Dynamics \textbf{3}(2), 023601 (2016)

\bibitem{Adachi2019}
S.~Adachi, T.~Schatteburg, A.~Humeniuk, R.~Mitri{\'c}, T.~Suzuki, Phys. Chem.
  Chem. Phys. \textbf{21}(26), 13902 (2019)

\bibitem{Serrano-Andres1993}
L.~Serrano-Andres, M.~Merchan, I.~Nebot-Gil, B.O. Roos, M.~Fulscher, J. Am.
  Chem. Soc. \textbf{115}(14), 6184 (1993)

\bibitem{Palmer1995}
M.H. Palmer, I.C. Walker, C.C. Ballard, M.F. Guest, Chem. Phys.
  \textbf{192}(2), 111 (1995)

\bibitem{Burcl2002}
R.~Burcl, R.D. Amos, N.C. Handy, Chem. Phys. Lett. \textbf{355}(1-2), 8 (2002)

\bibitem{Gromov2003}
E.~Gromov, A.~Trofimov, N.~Vitkovskaya, J.~Schirmer, H.~K{\"o}ppel, J. Chem.
  Phys. \textbf{119}(2), 737 (2003)

\bibitem{Gavrilov2008}
N.~Gavrilov, S.~Salzmann, C.M. Marian, Chem. Phys. \textbf{349}(1-3), 269
  (2008)

\bibitem{Stenrup2011}
M.~Stenrup, {\AA}.~Larson, Chem. Phys. \textbf{379}(1-3), 6 (2011)

\bibitem{Bearpark1994}
M.J. Bearpark, M.A. Robb, H.B. Schlegel, Chem. Phys. Lett. \textbf{223}(3), 269
  (1994)

\bibitem{Harabuchi2019}
Y.~Harabuchi, M.~Hatanaka, S.~Maeda, Chemical Physics Letters: X \textbf{2},
  100007 (2019)

\bibitem{Levine2006a}
B.G. Levine, C.~Ko, J.~Quenneville, T.J. Mart{\'I}nez, Mol. Phys.
  \textbf{104}(5-7), 1039 (2006)

\bibitem{Meng2008}
S.~Meng, E.~Kaxiras, J. Chem. Phys. \textbf{129}(5), 054110 (2008)

\bibitem{Meng2008a}
S.~Meng, J.~Ren, E.~Kaxiras, Nano Lett. \textbf{8}(10), 3266 (2008)

\end{thebibliography}
\end{document}